\documentclass[prl,reprint,showpacs,twocolumn,superscriptaddress,floatfix]{revtex4-1}
\usepackage{lipsum}

\usepackage{lineno}
\usepackage[dvipsnames]{xcolor}

\usepackage{bm}

\usepackage{graphicx}

\usepackage{upgreek}

\usepackage{natbib}

\usepackage{braket}

\usepackage{physics}
\usepackage[%
	breaklinks,%
	pdftex,%
	hyperfootnotes=true,%
	pdfpagelabels,%
	bookmarks,%
	pageanchor,%
]{hyperref}

\hypersetup{%
	colorlinks=true, linktocpage=true, pdfstartpage=1, pdfstartview=FitH, pdfborder={0 0 0},%
	breaklinks=true, pdfpagemode=UseNone, pageanchor=true, pdfpagemode=UseOutlines,%
	plainpages=false, bookmarksnumbered, bookmarksopen=true, bookmarksopenlevel=1,%
	hypertexnames=true, pdfhighlight=/O,
	urlcolor=Red!50!Sepia, linkcolor=NavyBlue, citecolor=RoyalBlue, %pagecolor=RoyalBlue,%
}

\graphicspath{{Pictures/}}

\newcommand{\ped}[1]{_\text{#1}}

\renewcommand{\epsilon}{\varepsilon}

\newcommand{\beq}{\begin{equation}}
\newcommand{\eneq}{\end{equation}}

\begin{document}
   
    \author{L.~M.~Cangemi}
    \affiliation{Dipartimento di Fisica ``E.~Pancini'', Universit\`a di Napoli ``Federico II'', Complesso di Monte S.~Angelo, via Cinthia, 80126 Napoli, Italy}
    \email{lorismaria.cangemi@unina.it}
    \affiliation{CNR-SPIN, c/o Complesso di Monte S. Angelo, via Cinthia - 80126 - Napoli, Italy}
    \author{M.~Carrega}
    \affiliation{CNR-SPIN,  Via  Dodecaneso  33,  16146  Genova, Italy}
    %\affiliation{NEST, Istituto Nanoscienze-CNR, I-56126 Pisa, Italy}
    \author{A.~De~Candia}
    \affiliation{Dipartimento di Fisica ``E.~Pancini'', Universit\`a di Napoli ``Federico II'', Complesso di Monte S.~Angelo, via Cinthia, 80126 Napoli, Italy}
    \affiliation{CNR-SPIN, c/o Complesso di Monte S. Angelo, via Cinthia - 80126 - Napoli, Italy}
    
    \affiliation{INFN, Sezione di Napoli, Complesso Universitario di Monte S. Angelo,
I-80126 Napoli, Italy}
    \author{V.~Cataudella}
    \affiliation{Dipartimento di Fisica ``E.~Pancini'', Universit\`a di Napoli ``Federico II'', Complesso di Monte S.~Angelo, via Cinthia, 80126 Napoli, Italy}
    \affiliation{CNR-SPIN, c/o Complesso di Monte S. Angelo, via Cinthia - 80126 - Napoli, Italy}
    \affiliation{INFN, Sezione di Napoli, Complesso Universitario di Monte S. Angelo,
I-80126 Napoli, Italy}
        \author{G.~De Filippis}
    \affiliation{Dipartimento di Fisica ``E.~Pancini'', Universit\`a di Napoli ``Federico II'', Complesso di Monte S.~Angelo, via Cinthia, 80126 Napoli, Italy}
    \affiliation{CNR-SPIN, c/o Complesso di Monte S. Angelo, via Cinthia - 80126 - Napoli, Italy}
    \affiliation{INFN, Sezione di Napoli, Complesso Universitario di Monte S. Angelo,
I-80126 Napoli, Italy}
\author{M.~Sassetti}
    \affiliation{Dipartimento di Fisica, Universit\`a di Genova, Via Dodecaneso 33, 16146 Genova, Italy} 
    \affiliation{CNR-SPIN,  Via  Dodecaneso  33,  16146  Genova, Italy}
    \author{G.~Benenti}
    \affiliation{Center for Nonlinear and Complex Systems, Dipartimento di Scienza e Alta Tecnologia, Universit\`a degli Studi dell'Insubria, via Valleggio 11, 22100 Como, Italy} 
    \affiliation{Istituto Nazionale di Fisica Nucleare, Sezione di Milano, via Celoria 16, 20133 Milano, Italy}
    \affiliation{NEST, Istituto Nanoscienze-CNR, I-56126 Pisa, Italy}

    \title{Optimal energy conversion through anti-adiabatic driving
\\breaking time-reversal symmetry}
    
    \date{\today}
    
    
    \begin{abstract}
Starting with Carnot engine, the ideal efficiency of a heat engine has been associated with quasi-static transformations
and vanishingly small output power.
Here, we exactly calculate the thermodynamic properties of a isothermal heat engine, 
in which the working medium is a periodically driven underdamped harmonic oscillator,
focusing instead on the opposite, anti-adiabatic limit, where the period of a cycle is the
fastest time scale in the problem.
We show that in that limit it is possible to approach the ideal
energy conversion efficiency $\eta=1$, with finite output power and 
vanishingly small relative power fluctuations. The simultaneous realization of 
all the three desiderata of a heat engine is possible thanks to the 
breaking of time-reversal symmetry. We also show that non-Markovian 
dynamics can further improve the power-efficiency trade-off. 	
    \end{abstract}
    
    \maketitle
    {\it Introduction.}---
Since its inception, the development of thermodynamics and its 
technological applications have been boosted by fundamental questions~\cite{Campisi:Noneq2,Seifert_2012:Review, kosloff13, Kurizki2015, Sothmann2015, Goold:ThermoInfo, Anders2016, BENENTI20171}.  
A key question is what are the ultimate bounds to the performance 
of heat engines~\cite{whitney, niedenzu, shiraishipre, CampisiPekolaFazio:HeatEngines, vischi, coherence18, bauerpre, skaller, brandnersaito,mitra, ito, gambetta20}. It is desirable that a heat engine operates close 
to the ideal efficiency 
(Carnot efficiency when the working medium exchanges heat with 
reservoirs at different temperatures, or unit efficiency 
for a isothermal engine),
delivers large power, and exhibits small power fluctuations~\cite{Holubec17, Pietzonka:Tradeoff}. 
Physical intuition tells us that the ideal efficiency can be obtained
only if the energy conversion process is reversible. 
As a thermodynamic reversible transformation is quasi-static, 
the thermodynamic cycle takes an infinite time, and therefore 
the output power vanishes. On the other hand, the second law of 
thermodynamics by itself does not forbid the possibility of 
achieving the ideal efficiency at finite power~\cite{Benenti:TR}. 
Several studies have shown that it is possible to come 
arbitrarily close to the ideal efficiency~\cite{Allahverdyan:Carnot,Shiraishi2015,Campisi:CriticalHeatEng,Konig2016,Polettini2017,Lee2017,Holubec17b,Holubec:Cycling}, 
with finite power. However, diverging power fluctuations, 
as in the case when the working medium is at the verge of a
phase transition~\cite{Campisi:CriticalHeatEng}, or 
the necessity of precisely engineering the scaling of model parameters~\cite{Holubec17b,Holubec:Cycling}, 
make such engines impractical~\cite{solon18, Holubec17,Holubec17b}.

Thermodynamic uncertainty relations (TURs)~\cite{uffin, Barato2015,gingrich2016,Seifert:annual,Horowitz:TUR, Brandner2018, Agarwalla:vioTUR2, liusegal19, timpanaro, Hasegawa2019, proesmans19, Potts2019, asegava, koyuk:boundper,Kheradsoud2019, Falasco2020, asegava} 
set a lower bound on the
time-integrated relative fluctuation of an arbitrary current,
which diverges when the dissipationless limit required for 
ideal efficiency is achieved.
The application of such relations to the ``work current'' (i.e., the
output power) leads,   
for steady-state heat engines with time-reversal symmetry (TRS), 
to a trade-off between efficiency, power,
and fluctuations~\cite{Pietzonka:Tradeoff}.
Consequently, approaching the ideal efficiency with 
finite power implies diverging fluctuations. On the other hand, such 
result does not apply for cyclic heat engines, for which a less restrictive trade-off
has been derived~\cite{koyuk:boundper} for overdamped Markovian dynamics.   
This interesting result raises the following questions: (i) Is it possible 
to approach the ideal efficiency at finite power and
finite (or even vanishing) relative power fluctuations 
in a purely dynamical model, without using the overdamped and Markov approximations?
(ii) Counterintuitive as it may be, 
is it possible to obtain such result 
far from
the quasi-static limit? Possibly in
the complementary, anti-adiabatic limit, where the period of a
cycle is the fastest time scale?

To address these questions, we consider the paradigmatic model where 
the working medium is a harmonic oscillator coupled to a thermal bath.
The oscillator's canonically conjugated variables are separately coupled to 
periodic drives. This system can act as a isothermal heat engine. 
This model has several advantages: 
(i) it can be exactly solved, also in the far-from-equilibrium regime
and for strong system-bath coupling, without resorting to the overdamped 
or other approximations;
(ii) it is possible to break time reversibility and to address 
all driving regimes, from the quasi-static to the anti-adiabatic one;
(iii) non-Markovian effects are naturally included and can be tuned 
by engineering the bath spectral density.
Here, we show that 
the ideal, unit efficiency for energy conversion is achieved both in 
the quasi-static and in the anti-adiabatic limits. 
While in the first case the output power
vanishes, in the latter we also obtain the other two desiderata of 
a heat engine, that is, finite power and vanishing relative 
power fluctuations. Non-Markovian effects can then further improve the 
power-efficiency trade-off while approaching the unit-efficiency limit. 
Finally, we clarify the necessity of breaking TRS 
to obtain the above results.

    {\it Driven harmonic isothermal heat engine.}--- We consider energy conversion process through a driven (quantum) resonator connected to a thermal reservoir, whose total Hamiltonian is $H(t)= H\ped{S}(t)+H\ped{R} + H\ped{SR}$. As working medium we consider a single harmonic oscillator, linearly coupled through its canonical degrees of freedom to time-periodic external fields driving it out-of equilibrium,
    \begin{equation}\label{eq:resonator}
    H\ped{S}(t)= \frac{p^2}{2m} + \frac{1}{2}m\omega^2_{0} x^2 - \epsilon_1(t) x -\epsilon_2(t) p~, 	  
    \end{equation}   
    where $m$ and $\omega_0$ are the mass and characteristic frequency of the oscillator, respectively (hereafter we set $\hbar=k_{{\rm B}}=1$).
    External drives are periodic functions, $\epsilon_{1,2}(t)=\epsilon_{1,2}(t+{\cal T})$ with period ${\cal T}=2\pi/\omega$.
%We underline that, as one drive is coupled through momentum, the system 
%is intrinsically underdamped, i.e. the overdamped approximation 
%does not even apply in the quasi-static limit $\omega\to 0$~\cite{supp:supp}.
    
The system is unavoidably coupled with its surrounding environment, which behaves as a thermal bath causing noise and dissipation. This can be described by the standard Caldeira-Leggett model~\cite{caldeira:caldeira-leggett,Weiss:open-quantum2}, in terms of an infinite set of harmonic oscillators,
    \begin{equation}\label{eq:bath}
    H\ped{R}+H\ped{SR} =  \sum_{k=1}^{\infty} \frac{P^2_k}{2 m_k} + \frac{m_k \omega^2_k}{2}\qty(X_k - \frac{c_k}{m_k \omega^2_k}{x})^2, 
    \end{equation}
    linearly coupled to the oscillator position ${x}$.
    \begin{figure}[ht]
    	\centering
    	\includegraphics[scale=0.8]{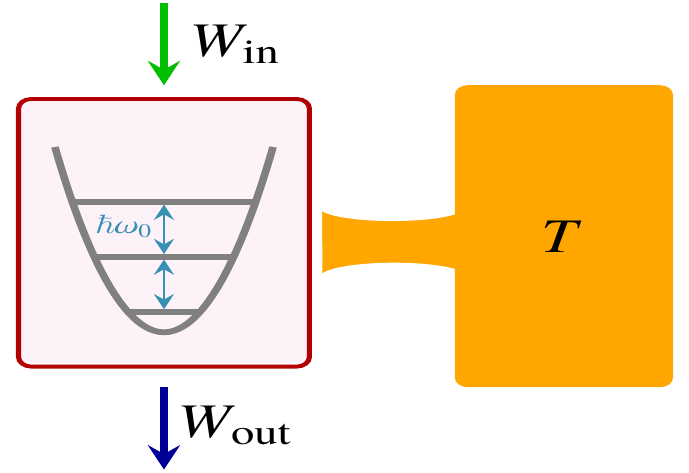}
    	\caption{Sketch of the driven harmonic isothermal heat engine.~A quantum resonator is periodically driven along two different channels.~It is connected to a thermal bath at fixed temperature $T$.~An amount of work per cycle is done on the system along a given channel ($W_{\text{in}}$), it is partially converted in output along the other ($W_{\text{out}}$) and partially dissipated into the bath.}
\label{fig:converter}
    \end{figure}
    The system can act as an isothermal heat engine~\cite{proesmans16, proesmans17, Carrega:wtow,Cangemi:violation}. 
%or work-to-work converter~\cite{proesmans16, proesmans17, Carrega:wtow,Cangemi:violation}.
    Physically, as sketched in Fig.~\ref{fig:converter}, when a given amount of work is put in an input channel, the system delivers part of it in output, while dissipating a given amount into the thermal reservoir at fixed temperature $T$.

The system described by Eq.~(\ref{eq:resonator}) can model a wide range of experimental realizations, covering both classical and quantum regimes. In the former case, one can consider an electronic RLC circuit coupled to both external time-dependent electric and magnetic fields. An ac voltage couples with the position variable (the charge), 
while the magnetic field 
%threading the circuit 
via electromagnetic induction 
couples with %the momentum variable 
the current.
%varying flux treading the magnetic field, by modifying the enclosing area, results in mechanical work, converting one form of energy into the other.
One can also think about a driven quantum LC circuit~\cite{IngoldNazarov:tunnel,Devoret1997QuantumFI,Girvin:cQED, espositof} implemented by superconducting circuit 
elements~\cite{Krantz:engqubits}.
The conjugate variables are the flux and charge variables $(\varphi,Q)$,
the capacitance $C$ plays the role 
of the mass $m$ and the oscillator frequency $\omega_0=1/\sqrt{LC}$, with 
$L$ the circuit inductance.
Here, one driving can be engineered, for example, by means of a capacitive coupling of an ac-voltage to the charge of the circuit~\cite{Girvin:cQED}. The other driving field can be obtained by using other inductive elements, which couple to the circuit via mutual inductance. The bath in Eq.~\eqref{eq:bath} phenomenologically accounts for dissipative effects on the LC circuit arising from the external circuitry impedance~\cite{Devoret1997QuantumFI}.

The equations of motion (EOM) read
    \begin{equation}\label{eq:eqmotion}
    \begin{gathered}
      \ev{\dot{x}(t)}= \frac{\ev{p(t)}}{m} -\epsilon_2(t), \\
         \ev{\ddot{x}(t)}+\int\limits_{-\infty}^t dt^{\prime} \gamma(t-t^{\prime}) \ev{\dot{x}(t^{\prime})} + \omega_0^2 \ev{x(t)} 
\\=\frac{\epsilon_1(t)}{m} -\dot{\epsilon}_2(t),
    \end{gathered}
    \end{equation}
    with $\ev{\dots }$ indicating the expectation value
over the initial state (at $t=-\infty$).
    Notice that Eq.~\eqref{eq:eqmotion}, which is valid both in 
the classical and in the quantum case, has the form of an underdamped Langevin equation in the presence of time-dependent forces.~The memory kernel $\gamma(t)$ describes friction and can be linked to the bath  spectral density $J(\omega)$
as $\gamma(t)=\frac{2}{\pi m }\theta(t)\int\limits_0^{+\infty}\mathrm{d}\omega J(\omega)\cos(\omega t)/ \omega$~\cite{supp:supp}.
    In the continuum limit, and assuming a Lorentzian high frequency cut-off $\omega\ped{c}$, $J(\omega)$ reads
    \begin{equation}\label{eq:spectral}
    J(\omega)=  m\gamma_s \bar{\omega}^{1-s} \frac{\omega^s}{1+ (\omega/\omega\ped{c})^2}.
    \end{equation}
    Here, $\gamma_s$ is the friction amplitude, $\bar{\omega}$ the characteristic bath frequency and the index $s$ distinguishes among different kinds of dissipation: Ohmic ($s=1$), sub-Ohmic ($s<1$) and super-Ohmic ($s>1$)~\cite{Weiss:open-quantum2}.
%, and in the case $s\neq 1$ they induce 
For $s\ne 1$, the system's dynamics is non-Markovian also
in the high frequency limit for the cut-off $\omega\ped{c}\to\infty$.
%At fixed spectral density, Eq.~\eqref{eq:eqmotion} fully describes the stochastic dynamics of the quantum resonator degrees of freedom, including non-Markovian memory effects linked to non-Ohmic dissipation.

The linearity of the EOM guarantees that the response to the external fields $\epsilon_{1/2}(t)$ can be computed exactly at any field strength.    
    The expectation values of $(x(t),p(t))$  can be written in terms of a generalized susceptibility matrix $\chi_{1/2,1/2}(t-t')=-i\theta(t-t')\ev{[x(t)/p(t),x(t')/p(t')]}_T$~\cite{note:kubo} as   
    \begin{equation}\label{eq: susceptibility}
    \begin{gathered}
	\ev{x(t)/p(t)}=-\sum_{j=1,2}\int_{-\infty}^{+\infty}dt'\chi_{1/2,j}(t-t')\epsilon_j(t').
    \end{gathered}
    \end{equation}
    Solving Eq.~\eqref{eq:eqmotion} in Fourier space, the analytic expressions for $\chi_{ij}(\omega)=\int\limits_{-\infty}^{+\infty}\mathrm{d}\tau\chi_{ij}(\tau)e^{i\omega\tau}$ follow~\cite{supp:supp}:
    \begin{align}\label{eq:chisol}
%    \begin{gathered}
    \chi_{11}(\omega)&=\frac{1}{m}\frac{1}{\omega^2-\omega_0^2+i\omega \gamma(\omega)},\nonumber\\
    \chi_{22}(\omega)&=- m + m^2\omega^2 \chi_{11}(\omega),\\
    \chi_{12}(\omega)&=-\chi_{21}(\omega)= i m \omega \chi_{11}(\omega),\nonumber
%    \end{gathered}
    \end{align}
    where $\gamma(\omega)=\int\limits_{0}^{+\infty}\gamma(\tau)e^{i\omega \tau}\mathrm{d}\tau$ is the frequency dependent damping~\cite{supp:supp}. It is worth to note that linearity imposes that $\chi_{ij}(\omega)$ do not depend on temperature.
   
    {\it Engine performance. }--- In the periodic steady state sustained by the drives, the total work per cycle can be defined as~\cite{jurgen}
${W}=\int_0^{{\cal T}} dt\, {\rm Tr}\left\{ \frac{\partial H(t)}{\partial t}\rho_{\rm tot}(t)\right\} =\int_0^{{\cal T}} dt\, {\rm Tr}\left\{ \frac{\partial H\ped{S}(t)}{\partial t}\rho_{\rm S}(t)\right\}$, with $\rho_{{\rm tot}}(t)$ and $\rho\ped{S}(t)$ the total and system density matrix at time $t$. The work contributions associated to the drives are identified as
\begin{equation}
{W}_{1/2}=-\int_0^{{\cal T}}dt \dot{\epsilon}_{1/2}(t)\ev{x(t)/p(t)}~,
\end{equation}
with mean power per cycle ${P}_{1/2}={W}_{1/2}/{{\cal T}}$. 
We consider the parameter regions where energy is exchanged 
between an input channel with ${P}_{{\rm in}}={P}_2>0$ 
and an output one with ${P}_{{\rm out}}=-{P}_1$ (${P}_1<0$)~\cite{note:note}, i.e. a given amount of work per cycle is converted from ${W}_{\text{in}}={W}_2$ to ${W}_{\text{out}}=-{W}_1$ with efficiency
\begin{equation}
\eta\equiv \frac{{W}_{\rm out}}{{W}_{\rm in}}=\frac{{P}_{{\rm out}}}{{P}_{{\rm in}}}~.
\end{equation}

For sake of definiteness, the external drives are chosen as $\epsilon_1(t)=\epsilon_1 \sin\omega t\mbox{, }\epsilon_2(t)=\epsilon_2 \cos(\omega t -\varphi)$, with $\varphi\in[0,2\pi)$ a possible phase shift. 
%Note that, due to the linearity of the EOM, signals with different (higher) harmonics will not give rise to finite response.
Exact espressions for the mean powers along the two channels can be computed starting from Eq.~\eqref{eq:eqmotion}, and can be compactly written as
\begin{equation}
\label{eq:sol}
{P}_i= \epsilon_i\sum_{j=1,2}\epsilon_j {\cal L}_{ij}~,
\end{equation}
where $\mathcal{L}_{ij}$ represent  the elements of the generalized Onsager matrix of the 
isothermal heat engine, which exactly describes the response of the working medium to the external fields.~In terms of susceptibilities $\chi_{ij}$, they read 
    \begin{equation}\label{eq:sol2}
    \begin{gathered}
    \mathcal{L}_{11}(\omega)=-\frac{\omega}{2}\Im[\chi_{11}(\omega)],\\
    \mathcal{L}_{22}(\omega)=-\frac{m^2\omega^3}{2}\Im[\chi_{11}(\omega)],\\
    \mathcal{L}_{12}(\omega,\varphi)=-\frac{m\omega^2}{2}\big(\sin\varphi\Re[\chi_{11}(\omega)] \\
+ \cos\varphi \Im[\chi_{11}(\omega)]\big),\\
    \mathcal{L}_{21}(\omega,\varphi)=\mathcal{L}_{12}(\omega,-\varphi),
    \end{gathered}
    \end{equation}
    and therefore they are temperature independent.
    It is worth to underline that the phase difference $\varphi$ controls the asymmetry of the Onsager matrix: for $\varphi\neq 0,\pi$, TRS is broken, with the device operating at maximum asymmetry, ${\cal L}_{12}=-{\cal L}_{21}$, for e.g. 
$\varphi=3\pi/2$. 
%

%Fluctuations of the output power can severely limit the precision of microscopic engines~\cite{Holubec17, solon18, Carrega:wtow}. These 
Fluctuations can be evaluated by computing the variance $D_{i}$ of $P_{i}$. Via fluctuation-dissipation theorem the time-averaged variance can be written as ${D}_i= \epsilon_i^2\omega\coth[\omega/(2T)]\mathcal{L}_{ii}$~\cite{supp:supp}, also valid for any field strength.
Contrary to mean powers and efficiency, which depend only on  ${\cal L}_{ij}$, quantum fluctuations explicitly depend on temperature, reaching only at high temperature their classical expression $\propto  T$~\cite{Weiss:open-quantum2}.

{\it Results.}--- We now characterize energy conversion performance,
focusing on the regimes where the efficiency $\eta$ is close to the ideal 
value $\eta=1$. 
To this end, for any given value of $\omega$ and $\varphi$,
we consider the maximum efficiency $\eta_{{\rm \scriptscriptstyle{ME}}}$, obtained by 
maximizing $\eta$ over the output amplitude $\epsilon_1$, for a fixed 
$\epsilon_2$.
All results discussed below are obtained considering the high frequency limit for the cut-off $\omega\ped{c}$.
Hereafter, we focus on the case with broken TRS, 
%($\varphi\neq 0,\pi$), 
showing that %(see also~\cite{supp:supp}) 
it is possible to achieve finite output power ${P}_{{\rm out,\scriptscriptstyle {ME}}}>0$ 
and vanishing relative output power fluctuations 
$\Sigma_{{\rm \scriptscriptstyle{ME}}}=\sqrt{{D}_{\rm out,\scriptscriptstyle{ME}}/{P}^2_{\rm out,\scriptscriptstyle{ME}}}$ 
with $\eta_{{\rm  \scriptscriptstyle{ME}}}\to 1$.
Explicit results are reported for the maximally asymmetric case 
%with maximum asymmetry, i.e. 
$\varphi=3\pi/2$; for any value of $\varphi$ breaking TRS, ${\cal L}_{12}=-{\cal L}_{21}$ 
is anyway recovered in the anti-adiabatic limit~\cite{supp:supp}.

\begin{figure}[ht]
    	\centering
    	\includegraphics[scale=0.60]{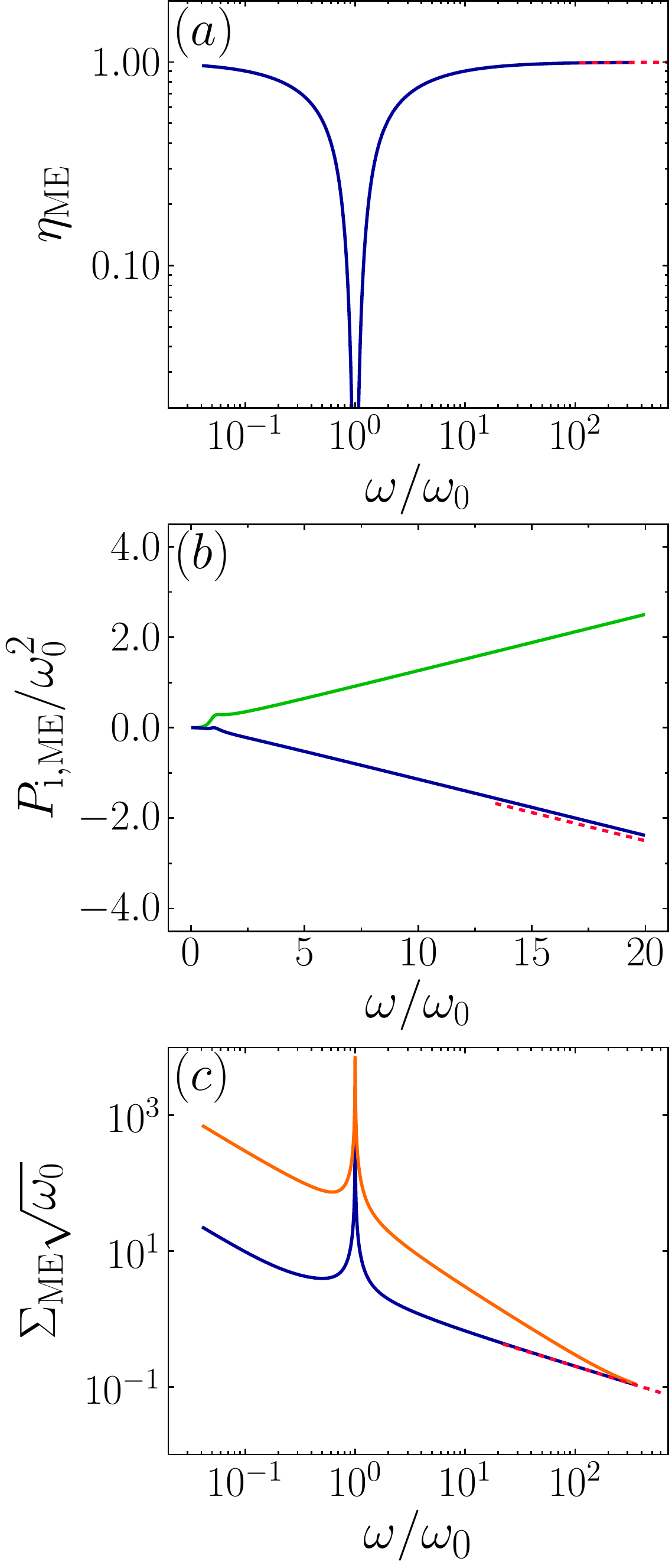}
    	\caption{Heat engine performance at ME as a function of the driving frequency $\omega$, for Ohmic friction $s=1$.
    	{Panel (a) shows the efficiency $\eta_{\rm \scriptscriptstyle{ME}}$, panel (b) the input (green solid line) and output (blue solid line) powers,} panel (c) the relative fluctuations at low temperature $T=0.1\,\omega_0$ (blue solid line) and high temperature 
$T=100\,\omega_0$ (orange solid line).
        Dashed lines correspond to the asymptotic power-law behavior for $\omega\gg\omega_0,\gamma_1$. All quantities are scaled in units of $\omega_0$. Other parameters are $m\epsilon_2^2=0.025\,\omega_0$, $\gamma_1=0.5\,\omega_0$, and $\bar{\omega}=\omega_0$.}
\label{fig:performance}
    \end{figure}

In Fig.~\ref{fig:performance} we show the efficiency $\eta_{{\rm \scriptscriptstyle{ME}}}$, the input and output power behaviors, and the relative power fluctuations 
at ME as a function of driving frequency $\omega$, in the case of Ohmic damping ($s=1$). As it is clear from Fig.~\ref{fig:performance}(a),
the ideal limit $\eta_{{\rm \scriptscriptstyle{ME}}}= 1$ is approached in two opposite regimes.
However, looking at Fig.~\ref{fig:performance}(b), the output power $P_{{\rm out, \scriptscriptstyle{ME}}}$ tends to vanish at small frequency $\omega\to 0$ (quasi-static limit). On the contrary,  it takes finite value in the anti-adiabatic driving regime {$\omega \gg \omega_0,\gamma_1$}.
More precisely, in the anti-adiabatic regime the output power has a linear scaling with frequency, 
$P_{{\rm out, \scriptscriptstyle{ME}}}=\frac{\epsilon^2_2 m}{2}\omega$, independently of the precise nature of dissipation (i.e., it does not depend on the parameter $s$).
Conversely, in the anti-adiabatic regime bath properties affect the scaling behavior of efficiency and fluctuations.
%(for a precise definition and details see~\cite{supp:supp}) 
%of other figures of merit. 
The efficiency approaches the 
ideal value as $\eta_{\rm \scriptscriptstyle{ME}}\to 1-2(\gamma_{\text{s}}/\bar{\omega})(\omega/\bar{\omega})^{s-2}$, with 
%$\overline{\gamma}_s=\gamma_s\overline{\omega}^{1-s}$ and 
$0<s<2$~\cite{supp:supp}.

Breaking of TRS (see discussion below) and anti-adiabatic driving 
allow us to approach the ideal efficiency with finite value of output power 
without affecting the engine precision.
Indeed, as shown in Fig.\ref{fig:performance}(c) the relative fluctuations $\Sigma_{\rm \scriptscriptstyle {ME}}$
are suppressed by increasing the frequency, at $\omega\gg \omega_0,\gamma_1$. The asymptotic scaling of
$\Sigma_{{\rm \scriptscriptstyle{ME}}}$, for generic dissipation, is given by 
$A\{\coth\qty[\omega/(2T)](\omega/\bar{\omega})^{(s-2)}\}^{1/2}$,
with the prefactor $A=(2\gamma_s/m \bar{\omega}\epsilon_2^2)^{1/2}$~\cite{supp:supp}, confirming the decrease of
$\Sigma_{{\rm \scriptscriptstyle{ME}}}$ with the increase of $\omega$, for $0<s<2$. In passing, we note that $\Sigma_{\rm \scriptscriptstyle{ME}}$ depends on temperature, implying lower fluctuations at low temperatures (quantum regime).
\begin{figure}[ht]
    	\centering
    	\includegraphics[scale=0.7]{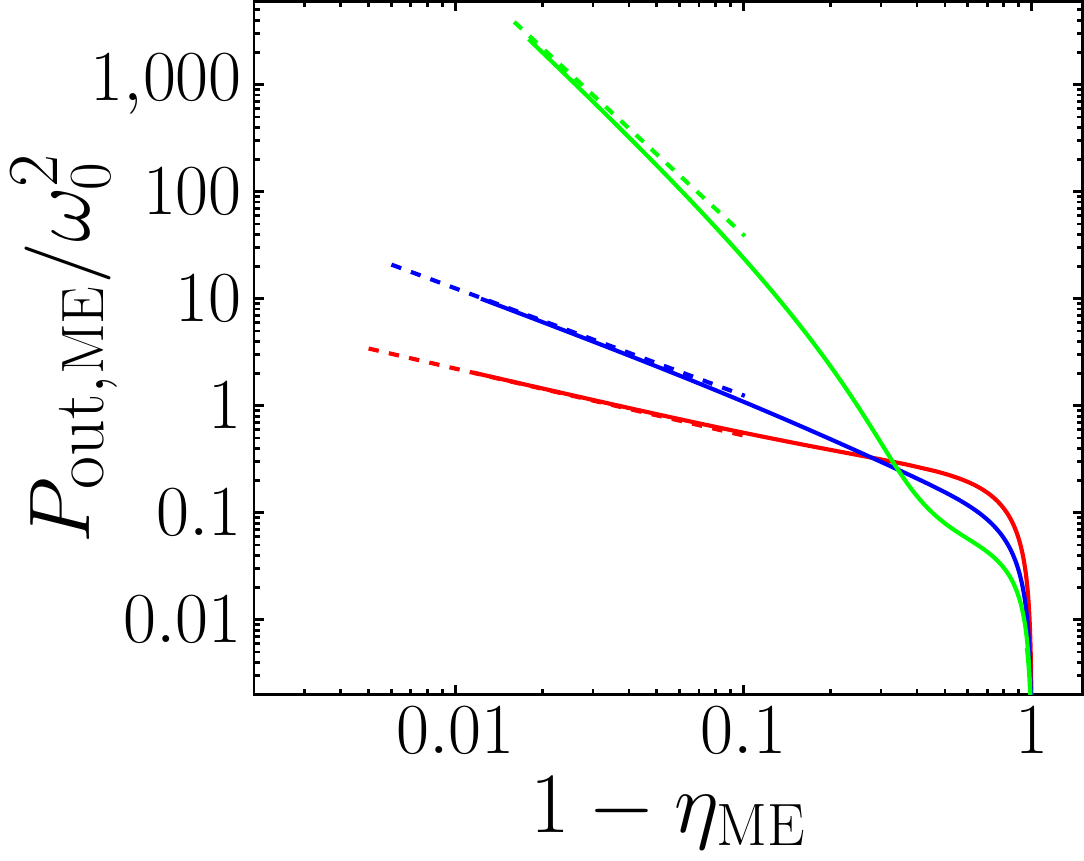}
    	\caption{Output power plotted versus $1-\eta_{{\rm \scriptscriptstyle{ME}}}$ at ME in the anti-adiabatic regime, for $s=0.4$ (red line), $s=1$ (blue line),
and $s=1.6$ (green line). 
The dashed lines correspond to the scaling $P_{\rm out,\scriptscriptstyle{ME}}\propto (1-\eta_{\rm \scriptscriptstyle{ME}})^{1/(s-2)}$.  Other parameter values are as in Fig.\ref{fig:performance}.}\label{fig:Pvseta}
    \end{figure}

To quantify engine performance, it is interesting to look at the scaling property of the output power while approaching the ideal limit $\eta_{\rm \scriptscriptstyle{ME}} \to 1$,
in the anti-adiabatic regime~\cite{supp:supp}. 
This is depicted in Fig.~\ref{fig:Pvseta}. Here, $P_{{\rm out,\scriptscriptstyle{ME}}}$ versus $1-\eta_{\rm \scriptscriptstyle{ME}}$ shows different power-law behaviors for different values of $s$, at fixed damping strength.
Non-Ohmic environment, implying memory effects, qualitatively  changes the scaling of power versus efficiency: $P_{\rm out,\scriptscriptstyle{ME}}\propto (1-\eta_{\rm \scriptscriptstyle{ME}})^{1/(s-2)}$.
Since the output power does not depend on $s$, 
it follows that 
sub-Ohmic dissipation ($s<1$) allows the conversion efficiency $\eta_{\rm \scriptscriptstyle{ME}}$ to get closer to the unit-efficiency limit, with respect to the other cases.
%Differently from other recent models of 
%heat engine, where a linear scaling $P_{{\rm out}}\propto 1-\eta$ is found, in our case the power-law dependence is non-linear 
%in the non-Ohmic case, 
%with the value of the exponent linked to the dissipation mechanism.
%Recently, in the context of thermodynamics uncertainty relation (TUR), a tradeoff between entropy production rate and relative fluctuations of the output power in steady-state engines has been proved~\cite{}.
%This is usually quantified by the parameter ${\cal Q}= (1/\eta -1) P_{{\rm out}} \Sigma^2 /T$~\cite{Pietzonka:Tradeoff}.
%Several bounds have been introduced for ${\cal Q}$~\cite{}, considering static driving\cite{} or obeying overdamped dynamics\cite{Koyuk:BoundPer}. As shown in \cite{supp:supp}, the model discussed here violates these bounds\cite{}, since it is in the fast driving response\cite{} and it obeys underdamped dynamics\cite{}.
Note that the bath properties drastically affect the entropy production rate $\sigma=(P_{\rm in}-P_{\rm out})/T$.
Since $\sigma\propto \omega^{s-1}$ for $\omega\to\infty$, vanishing entropy production rate is achieved in the sub-Ohmic regime. 

We mention that our model
%, intrinsically obeying underdamped dynamics, 
violates in a broad range of parameter values 
the TUR bound~\cite{koyuk:boundper}, 
derived for periodically driven, overdamped Markovian systems~\cite{supp:supp}. 
In particular, while for our model in the anti-adiabatic regime such bound predicts ${\cal Q}_{\rm \scriptscriptstyle{ME}}\equiv \sigma_{\rm \scriptscriptstyle{ME}}\Sigma_{\rm \scriptscriptstyle{ME}}^2\ge 2$,
we obtain ${\cal Q}_{\rm \scriptscriptstyle{ME}}\to 0$ when $\omega\to\infty$~\cite{supp:supp}.

Finally, we comment on the role played by TRS breaking.
Provided that linear response holds, a general trade-off between 
efficiency, power, and relative fluctuations can be 
derived following Ref.~\cite{Brandner:unifiedTUR}. 
For a isothermal heat engine we obtain
\begin{equation}
\label{eq:bound}
{\cal Q}
= \sigma\Sigma^2
=\left(\frac{1}{\eta} -1\right) 
\frac{P_{{\rm out}}}{T} {\Sigma^2} \geq {\cal B}\equiv \frac{2}{1 + S_L^2}~,
\end{equation}
where 
%$\sigma=(P_{\rm in}-P_{\rm out})/T$ is the entropy production rate and
$S_L$ is related to the asymmetry of the Onsager matrix~\cite{Brandner:unifiedTUR}. In particular, for systems with TRS $S_L=0$ and one recovers the 
bound of Ref.~\cite{Pietzonka:Tradeoff}, which forbids ideal efficiency at finite power, 
unless fluctuations diverge.
Our model, although not restricted to small values of the external driving strength, is written in terms of generalized Onsager coefficients and therefore satisfies the bound \eqref{eq:bound}~\cite{supp:supp}. 
%One may thus wonder how it is possible to get optimal performance without violating the bound of Eq.\eqref{eq:bound}. 
Breaking TRS implies a diverging asymmetry coefficient $S_L$ in the anti-adiabatic regime, i.e. ${\cal B}\to 0$. Therefore, in this limit the trade-off parameter ${\cal Q}_{\rm \scriptscriptstyle{ME}}\geq 0$, that is, it becomes irrelevant  
for the engine performance. 

{\it Conclusions.}---   	
By modeling a isothermal heat engine as a driven harmonic oscillator coupled to a 
thermal bath, we show that in the anti-adiabatic regime it is possible to achieve 
the ideal energy conversion, with simultaneous finite, and precise, output power.
Essential ingredients for our results are 
the breaking of TRS and the fact that the exact underdamped 
dynamics of the working medium is considered. 
We stress that in the opposite, quasi-static limit the above 
desirable features cannot be jointly observed, since the output power 
vanishes when the ideal, unit efficiency is approached. 

It would be interesting to investigate the exact underdamped 
dynamics in the case of a harmonic
oscillator coupled to two heat baths at different temperatures~\cite{saitosei, delcampo, pomu, miller,  bibek, Kadijani2020}.
In particular, to verify if also in this case it is possible to approach 
in the anti-adiabatic regime the ideal, Carnot limit at finite 
power and with finite or even vanishing relative fluctuations. 
Further natural generalizations of our model could be obtained
by considering engines with a more complex working medium, like coupled oscillators
or qubit-cavity systems.

We acknowledge G. Strini, F. Giazotto, S. Gasparinetti for useful discussions.
A.d.C. acknowledges financial support of the MIUR PRIN 2017WZFTZP
''Stochastic forecasting in complex systems''.

\bibliography{paper_armonic_v6}   

\begin{thebibliography}{70}
\bibitem [{not({\natexlab{a}})}]{note:kubo}%
  \BibitemOpen
  \href@noop {} {}\bibinfo {note} {$\langle ... \rangle_T$ denotes averaging
  over the thermal equilibrium state $\rho_T=\exp(-H_0/T)$, where $H_0$ is the
  overall Hamiltonian, excluding the driving terms, with $x$ and $p$ evolving
  with respect to $H_0$.}\BibitemShut {Stop}%
\bibitem [{not({\natexlab{b}})}]{note:note}%
  \BibitemOpen
  \href@noop {} {}\bibinfo {note} {If $P_1 > 0$, $P_2 < 0$ then
  $P_{\text{out}}=-P_2$, $P_{\text{in}}=P_1$.}\BibitemShut {Stop}%
\end{thebibliography}
         
\end{document}

\title{Supplemental materials for ''Optimal energy conversion through anti-adiabatic driving breaking time-reversal simmetry''}
    
\author{L.~M.~Cangemi}
    \affiliation{Dipartimento di Fisica ''E.~Pancini'', Universit\`a di Napoli ''Federico II'', Complesso di Monte S.~Angelo, via Cinthia, 80126 Napoli, Italy}
    \email{lorismaria.cangemi@unina.it}
    \affiliation{CNR-SPIN, c/o Complesso di Monte S. Angelo, via Cinthia - 80126 - Napoli, Italy}
    \author{M.~Carrega}
    \affiliation{CNR-SPIN,  Via  Dodecaneso  33,  16146  Genova, Italy}
    \author{A.~De~Candia}
    \affiliation{Dipartimento di Fisica ''E.~Pancini'', Universit\`a di Napoli ''Federico II'', Complesso di Monte S.~Angelo, via Cinthia, 80126 Napoli, Italy}
    \affiliation{CNR-SPIN, c/o Complesso di Monte S. Angelo, via Cinthia - 80126 - Napoli, Italy}
    \affiliation{INFN, Sezione di Napoli, Complesso Universitario di Monte S. Angelo,
I-80126 Napoli, Italy}
    
    \author{V.~Cataudella}
    \affiliation{Dipartimento di Fisica ''E.~Pancini'', Universit\`a di Napoli ''Federico II'', Complesso di Monte S.~Angelo, via Cinthia, 80126 Napoli, Italy}
    \affiliation{CNR-SPIN, c/o Complesso di Monte S. Angelo, via Cinthia - 80126 - Napoli, Italy}
\affiliation{INFN, Sezione di Napoli, Complesso Universitario di Monte S. Angelo,
I-80126 Napoli, Italy}
    
    \author{G.~De Filippis}
    \affiliation{Dipartimento di Fisica ''E.~Pancini'', Universit\`a di Napoli ''Federico II'', Complesso di Monte S.~Angelo, via Cinthia, 80126 Napoli, Italy}
    \affiliation{CNR-SPIN, c/o Complesso di Monte S. Angelo, via Cinthia - 80126 - Napoli, Italy}
    \affiliation{INFN, Sezione di Napoli, Complesso Universitario di Monte S. Angelo,
I-80126 Napoli, Italy}
    
    \author{M.~Sassetti}
    \affiliation{Dipartimento di Fisica, Universit\`a di Genova, Via Dodecaneso 33, 16146 Genova, Italy} 
    \affiliation{CNR-SPIN,  Via  Dodecaneso  33,  16146  Genova, Italy}
    \author{G.~Benenti}
    \affiliation{Center for Nonlinear and Complex Systems, Dipartimento di Scienza e Alta Tecnologia, Universit\`a degli Studi dell'Insubria, via Valleggio 11, 22100 Como, Italy} 
    \affiliation{Istituto Nazionale di Fisica Nucleare, Sezione di Milano, via Celoria 16, 20133 Milano, Italy}
    \affiliation{NEST, Istituto Nanoscienze-CNR, I-56126 Pisa, Italy}
   % \affiliation{NEST, Istituto Nanoscienze-CNR and Scuola Normale Superiore, Piazza S. Silvestro 12, I-56127 Pisa, Italy}
\maketitle
\section{Solution of the equation of motion}
We start from the Hamiltonian reported in the main text (recall that we set  $\hbar=1$ and $k_{B}=1$),
\begin{equation}\label{eq:hamtotsupp}
H = H_{{\rm S}}(t) + H_{{\rm R}} + H_{{\rm SR}},
\end{equation}
where 
\begin{equation}\label{eq:hamssupp}
H_{{\rm S}}(t)= \frac{p^2}{2m} + \frac{1}{2}m\omega^2_{0} x^2 - \epsilon_1(t) x -\epsilon_2(t) p~, 	  
\end{equation}
is the driven resonator Hamiltonian,
\begin{equation}\label{eq:bathsupp}
H_{{\rm R}}=  \sum_{k=1}^{\infty} \qty[\frac{P^2_k}{2 m_k} + \frac{m_k \omega^2_k X^2_k}{2}], 
\end{equation}
is the collection of harmonic oscillators describing the reservoir and  
\begin{equation}\label{eq:intsupp}
H_{{\rm SR}} =  -x\sum_{k=1}^{\infty} c_kX_k +x^2 \sum_{k=1}^{\infty}\frac{c^2_k}{2m_k \omega^2_k}, 
\end{equation}
 represents the coupling, bilinear in the position operators of system and reservoir. Notice that the last  term in Eq.(\ref{eq:intsupp}) contains only an operator acting in the system Hilbert space
but it depends on the coupling constants $c_k$. The physical reason for the inclusion
of this term is to avoid the potential renormalization introduced by the first term in Eq.(\ref{eq:intsupp}) \cite{Weiss:open-quantum2}.

The driven dissipative model, in Eq.~(\ref{eq:hamtotsupp}), allows for an analytical solution that holds for any value of the field amplitudes.
We now derive the equations of motion (EOM) in the presence of external driving fields $\epsilon_{1/2}(t)$.
From Eq.~\eqref{eq:hamtotsupp}, the time evolutions for the resonator operators $(x(t),p(t))$, in the Heisenberg picture, read    
\begin{eqnarray}\label{eq:EOM}
\dot{x}(t)&\!\!=\!\!& \frac{p(t)}{m} -\epsilon_2(t),\nonumber\\
\dot{p}(t)&\!\!=\!\!&-m\omega^2_{0} x(t)+\!\!\sum_{k=1}^{\infty} c_k \qty(X_k-\frac{c_k}{m_k \omega^2_k} x(t))\!+\!\epsilon_1(t), 
\end{eqnarray}
similarly, the  time evolutions for the oscillator operators $(X_k(t),P_k(t))$ are
\begin{eqnarray}\label{eq:EOM1}
\dot{X_k}(t)&=& \frac{P_k(t)}{m_k},\nonumber\\
\dot{P_k}(t)&=&- m_k\omega^2_{k} X_k(t) + c_k x(t).
\end{eqnarray}
Solving the EOM for the bath degrees of freedom as a function of the operator $x(t)$ and  substituting into Eq.~\eqref{eq:EOM} one obtains the so-called  generalized quantum Langevin equation \cite{Weiss:open-quantum2}
\begin{eqnarray}\label{eq:diffqoper}
&&\ddot{x}(t) + \int_{t_0}^{+\infty}\!\!\!\!\mathrm{d}t' \gamma(t-t')\dot{x}(t') + \omega_0^2 x(t)=-\gamma(t-t_0)x(t_0)\nonumber\\
&&+\frac{\xi(t)+\epsilon_1(t)}{m} -\dot{\epsilon}_2(t),
\end{eqnarray}
with $t_0\to-\infty$ the initial time. The function 
\begin{equation}\label{eq:gamma}
\gamma(t)=\frac{\theta(t)}{m}\sum_{k=1}^{\infty}\frac{c_k^2}{m_k\omega_k^2}\cos(\omega_k t),
\end{equation}
represents the memory damping kernel and 
\begin{equation}
\label{randomF}
\!\!\xi(t)=\!\!\sum_{k=1}^{\infty}c_k\left[X_k(t_0)\cos\omega_k (t-t_0)+\frac{P_k(t_0)}{m_k\omega_k}\sin\omega_k (t-t_0)\right],
\end{equation}
is the operator-valued fluctuating force which depends explicitly on the initial conditions of the bath position/momentum operators $X_k(t_0)$ and $P_k(t_0)$.  Notice that the inhomogeneous term  $\gamma(t-t_0)x(t_0)$ on the l.h.s. of Eq.(\ref{eq:diffqoper}) is a typical transient contribution and in our case, with $t_0\to -\infty$, for times $t>0$  decays to zero. 

In order to specify the reduced system dynamics it is now necessary  to define the initial condition of the density matrix, which will also fix the statistical properties of the quantum noise. 
The usual choice, which allows  to consider  $\xi(t)$ as a stochastic force with zero average, is a factorized initial 
preparation $\rho_{\rm tot}(t_0) = \rho_S(t_0)\otimes \rho_R(t_0)$\\
given by the product between  the initial system density $\rho_S(t_0)$ and the density matrix of the bath alone, which is assumed in thermal equilibrium 
$\rho_R(t_0)=\exp(-H_{\rm R}/T)/\Tr{\exp(-H_{\rm R}/T)}$.
With this choice we can evaluate the averages over the initial total density matrix of time dependent operators associated both to the bath and to the system. In the following the notation  $\langle A\rangle\equiv\Tr{\rho_{\rm tot}(t_0)A}$ will represent the quantum average of a generic operator $A$. It is easy to see that $\langle\xi(t)\rangle=0$, and consequently from Eq. (\ref{eq:diffqoper}) the following EOM for the average position $\langle x(t)\rangle$ ($t_0\to-\infty)$
\begin{equation}\label{eq:diffq}
\langle \ddot{x}(t)\rangle + \int_{-\infty}^{+\infty} \mathrm{d}t' \gamma(t-t')\langle \dot{x}(t')\rangle + \omega_0^2 \langle x(t)\rangle =\frac{\epsilon_1(t)}{m} -\dot{\epsilon}_2(t).
\end{equation}
It is worth to notice that this equation, derived starting from quantum operators, is equivalent to the classical Langevin equation. The possible quantum features are eventually inside the properties of the reservoir  treated as a quantum bath. 
Within the reduced description of the system alone, all quantities characterizing the environment are expressed in terms of the spectral density of the bath 
\begin{equation}
J(\omega)=\frac{\pi}{2}\sum_{k=1}^\infty \frac{c_k^2}{m_k\omega_k}\delta(\omega-\omega_k).
\end{equation}
For example the damping kernel $\gamma(t)$ defined in Eq.(\ref{eq:gamma}) can be written as 
\begin{equation}\label{eq:gammt}
\gamma(t)=\frac{2}{\pi m}\theta(t)\int_0^{\infty}\mathrm{d}\omega \frac{J(\omega)}{\omega}\cos(\omega t).
\end{equation}
Also the noise correlation function of the fluctuating force $\xi(t)$ becomes  $\langle\xi(t)\xi(t')\rangle=\langle\xi(t-t')\xi(0)\rangle$ with \cite{Weiss:open-quantum2}
\begin{equation}
\label{xicorr}
\langle\xi(t)\xi(0)\rangle=\hbar\!\!\int_0^{\infty}\!\!\frac{\mathrm{d}\omega}{\pi} 
J(\omega)\!\!\left[\coth(\frac{\hbar\omega}{2T})\cos(\omega t)-i\sin(\omega t)\right].
\end{equation}
In the last expression we have reintroduced the $\hbar$ factor in order to emphasize  the quantum character of the stochastic force correlators which in the classical case would corresponds to the limit $\hbar\to 0$, i.e. $\langle\xi(t)\xi(0)\rangle=m\gamma(t) T$.

To determine the time evolution of the system we now need to specify the frequency behavior of $J(\omega)$, and in particular its low frequency scaling, which determines the long time dynamics. We thus consider the realistic class of reservoirs with a power law behavior $J(\omega)\propto \omega^s$ at  low frequency, here the index $s$ distinguishes between different dissipative influence: $s=1$ corresponds to an Ohmic bath, $s<1$ sub-Ohmic behaviour and $s>1$ super-Ohmic behavior. In the following, we will focus on the range $0<s<2$. 
In order to describe a realistic heat bath, a cut off for the spectral density at high frequencies have to be considered. As done in the main text,  we  choose an algebraic lorentzian cut-off, with the reservoir  described in the continuum limit by
\begin{equation}\label{ap:spectral}
J(\omega)=  m\gamma_{\rm s} \bar{\omega}^{1-s} \frac{\omega^s}{1+ (\omega/\omega_{\rm{c}})^2},
\end{equation}
where $\gamma_{\rm s}$ is the friction amplitude, $\bar{\omega}$ is a characteristic frequency of the bath and $\omega_c$ is the largest  frequency playing the role of the cut-off.

We  recall that we are interested in the dynamics at times $t>0$. In this case the transient homogeneous part, which depend on the initial condition averages $\langle x(-\infty)\rangle$, $\langle p(-\infty)\rangle)$, will decay and vanish in the interval $(-\infty,0)$.
We are then left to find only the inhomogeneous solution driven by the external fields $\epsilon_{1/2}$. This solution can be conveniently obtained transforming Eq. \eqref{eq:diffq} in the frequency domain. Defining the  Fourier transform 
$
{f}(\omega)=\int_{-\infty}^{+\infty}\mathrm{d}t e^{i\omega t}f(t)
$
we have
\begin{equation}\label{eq:Foureom}
-\omega^2\ev{{x}(\omega)} - i \omega {\gamma}(\omega)\ev{ {x}(\omega)} + \omega_0^2 \ev{{x}(\omega)} = \frac{{\epsilon}_1(\omega)}{m} + i \omega {\epsilon}_2(\omega). 
\end{equation}
From Eqs. \eqref{eq:Foureom} and \eqref{eq:EOM} we obtain the exact solutions 
\begin{equation}\label{eq:solutions}
\begin{aligned}
\langle {x}(\omega)\rangle &=-{\chi}_{11}(\omega)[{\epsilon}_1(\omega) + i m \omega {\epsilon}_2(\omega)],\\
\langle {p}(\omega)\rangle &=  im{\chi}_{11}(\omega)[{\epsilon}_1(\omega) + i m \omega {\epsilon}_2(\omega)] + m {\epsilon}_2(\omega). 
\end{aligned}
\end{equation}
written in term of  the dynamical susceptibility
\begin{equation}
\label{chi11}
{\chi}_{11}(\omega)=\frac{1}{m}\frac{1}{\omega^2-\omega_0^2+i\omega{\gamma}(\omega)}.
\end{equation}
 In the time domain this latter function corresponds to the retarded response function
\begin{equation}
\chi_{11}(t-t')=-i\theta(t-t')\langle [x(t),x(t')]\rangle_T,
\end{equation}
where the average $\langle...\rangle_T$ is over  the thermal  equilibrium density matrix
$\rho_T=\exp(-H_{0}/T)/\Tr{\exp(-H_{0}/T)}$, with $H_0$  the overall Hamiltonian in Eq.(\ref{eq:hamtotsupp}), excluding the driving terms $\epsilon_{1/2}(t)$, and the time evolution of $x(t)$ is with respect to $H_0$.
The solutions (\ref{eq:solutions}) can be rewritten in a more compact way introducing  additional response functions $\chi_{ij}(\omega)$ defined as 
\begin{eqnarray}\label{eq:Chis}
 {\chi}_{12}(\omega)&=&im \omega  {\chi}_{11}(\omega),\nonumber\\
{\chi}_{21}(\omega)&=&-im \omega {\chi}_{11}(\omega),\nonumber\\
{\chi}_{22}(\omega)&=& - m +m^2\omega^2{\chi}_{11}(\omega).
\end{eqnarray}
These correspond to all possible additional correlators with $x$ and $p$ operators, namely
\begin{equation}\label{eq:commu}
\begin{gathered}
\chi_{12}(t-t')=-i \theta(t-t')\langle [x(t),p(t')]\rangle_T,\\
\chi_{21}(t-t')=-i \theta(t-t')\langle [p(t),x(t')]\rangle_T,\\
\chi_{22}(t-t')=-i \theta(t-t')\langle [p(t),p(t')]\rangle_T.
\end{gathered}
\end{equation}
Using these correlators we can write the solutions as
\begin{equation}\label{eq:linresp2}
\begin{gathered}
\ev{{x}(\omega)} = - \chi_{11}(\omega){\epsilon}_1(\omega) - {\chi}_{12}(\omega){\epsilon}_2(\omega),\\
\ev{{p}(\omega)} = - {\chi}_{21}(\omega){\epsilon}_1(\omega)-{\chi}_{22}(\omega){\epsilon}_2(\omega).
\end{gathered}
\end{equation}
%We conclude now this part adding few additional comments.  
Before closing this part, some comments are in order.
(i) The linearity of the EOM \eqref{eq:diffq} implies an exact solution  linear in the external fields $\epsilon_{1/2}$ without higher order contributions, in addition the retarded correlators $\chi_{i,j}(\omega)$ do not depend on temperature and are equal to their classical counterparts.

(ii) As quoted in the main part, the solution in Eq. (\ref{eq:linresp2}) can be also expressed in the time domain as a convolution
\begin{eqnarray}\label{xtime}
\langle x(t)\rangle &=&-\sum_{j=1}^{2}\int_{-\infty}^{+\infty}\mathrm{d}t' \chi_{1j}(t-t') \epsilon_j(t'),\nonumber\\
\langle p(t)\rangle &=&-\sum_{j=1}^{2}\int_{-\infty}^{+\infty}\mathrm{d}t' \chi_{2j}(t-t') \epsilon_j(t').
\end{eqnarray}

(iii) The memory-friction kernel $\gamma(t-t')$, present in the Langevin equation, is completely determined via its Fourier transform $\gamma(\omega)$ which,  using the Lorenzian shape of the spectral density in Eq.~\eqref{ap:spectral},  can be written as 
\begin{equation}
{\gamma}(\omega)=\frac{2{\gamma}_{\rm s}}{\pi}\int_{0}^{\infty} \mathrm{d}\omega_1 \frac{(\omega_1/\bar{\omega})^{s-1}}{1+(\omega_1/\omega_c)^2}\int_{0}^{\infty}\mathrm{d}t\cos(\omega_1 t)e^{i\omega t}.
\end{equation}

Employing now the well-known identity $\int_{-\infty}^{+\infty}dx~e^{-i\omega x}\theta(x)=\pi \delta(\omega) -i \text{P.V.}( 1/\omega)$, 
with the last term denoting the principal value, we obtain an explicit result  for the  real and imaginary parts of ${\gamma}(\omega)={\gamma}'(\omega) + i {\gamma}''(\omega)$
\begin{eqnarray}
{\gamma}'(\omega)\!\!&=&\!\!{\gamma}_s\abs{\frac{\omega}{\bar{\omega}}}^{s-1}
\frac{1}{1+(\omega/\omega_c)^2},\\
{\gamma}''(\omega)\!\!&=&\!\!\frac{{\gamma}_s \text{sgn}{\omega}}{1+(\omega/\omega_c)^2}
\abs{\frac{\omega}{\bar{\omega}}}^{s-1}\qty[\cot(\pi s/2)+\abs{\frac{\omega}{\omega_c}}^{2-s}
\!\!\!\!\frac{1}{\sin(\pi s/2)}].\nonumber
\end{eqnarray}
Below we quote the behavior of $\gamma(\omega)$ in the scaling limit,~\ie~$\omega\ll\omega_c$ which is often used in the main text.  We have
\begin{eqnarray}
\label{scaling}
{\gamma}'(\omega)&=&\gamma_{\rm s} \abs{\frac{\omega}{\bar{\omega}}}^{s-1},\nonumber\\
{\gamma}''(\omega)&=&\gamma_{\rm s}\cot (\pi s/2) {\rm sgn}(\omega)\abs{\frac{\omega}{\bar{\omega}}}^{s-1},
\end{eqnarray}
valid for $0< s< 2$. 

\section{Average powers and their fluctuations}

The power contributions along the two different channels can be written as
\begin{equation}
\begin{gathered}
P_1(t)=-\dot{\epsilon}_1(t)\ev{x(t)},\\
P_2(t)=-\dot{\epsilon}_2(t)\ev{p(t)}.
\end{gathered}
\end{equation}
Using Eq.~\eqref{xtime}, the powers become 
\begin{equation}\label{eq:powers}
P_{i}(t)=\dot{\epsilon}_i(t)\sum_{j=1}^{2}\int_{-\infty}^{+\infty}\mathrm{d}\tau \chi_{ij}(\tau) \epsilon_j(t-\tau).
\end{equation}
Hereafter, we fix the fields $(\epsilon_1(t),\epsilon_2(t))$, as done in the main text: We choose monochromatic functions with equal frequency $\omega$ with a phase shift $\varphi$
\begin{eqnarray}\label{eq:fields}
\epsilon_1(t)&=& \epsilon_1 \sin(\omega t),\nonumber\\
\epsilon_2(t)&=&\epsilon_2\cos(\omega t-\varphi).
\end{eqnarray}
With this choice it is easy to see that the powers $P_i(t)$  are periodic functions of time $t$, with period $\mathcal{T}=2\pi/\omega$.~We can thus compute their average over the period $\mathcal{T}$ defined as 
\begin{equation}
{P}_i=\frac{1}{\mathcal{T}}\int_{\bar t}^{\mathcal{T}+{\bar t}}\mathrm{d}t P_i (t),
\end{equation}
with $\bar t$ a generic positive time.
Notice that, due to the bilinear form  in the external fields  (\ref{eq:powers}), signals with  different frequencies will not give rise to a finite average power. This explains why we fixed the same frequency $\omega$ in $\epsilon_1$ and $\epsilon_2$.
~By substituting Eq.~\eqref{eq:fields} into Eq.~\eqref{eq:powers} and averaging over $\mathcal{T}$ we obtain
\begin{equation}\label{eq:sol}
{P}_i=\epsilon_i\sum_{j=1,2}\epsilon_j {\cal L}_{ij}(\omega),~
\end{equation}
where ${\mathcal L}_{ij}(\omega)$ are generalized Onsager coefficients. After straightforward manipulations, they can be written in terms of the retarded response functions \eqref{chi11} as follows
\begin{eqnarray}\label{eq:Onsager}
\!\!\!\!&&\mathcal{L}_{11}(\omega)=-\frac{\omega}{2}\Im[\chi_{11}(\omega)],\nonumber\\
\!\!\!\!&&\mathcal{L}_{22}(\omega)=-\frac{m^2\omega^3}{2}\Im[\chi_{11}(\omega)],\nonumber\\
\!\!\!\!&&\mathcal{L}_{12}(\omega,\varphi)=-\frac{m\omega^2}{2}\big[\sin\varphi\Re[\chi_{11}(\omega)]+\cos\varphi \Im[\chi_{11}(\omega)]\big],\nonumber\\
\!\!\!\!&&\mathcal{L}_{21}(\omega,\varphi)=\mathcal{L}_{12}(\omega,-\varphi).
\end{eqnarray}  
Eqs.~\eqref{eq:sol}, \eqref{eq:Onsager} allow us to compute the mean powers of the isothermal engine for given values of the driving frequency $\omega$, phase difference $\varphi$ and field amplitudes $(\epsilon_1,\epsilon_2)$.

Due to the presence of dissipation, the mean powers in Eq.~\eqref{eq:sol} undergo fluctuations, which would affect the engine performance.
Fluctuations, during the whole time interval $t-t_0$, are described in terms of the power autocorrelation function. We thus introduce the deviation power operators
\begin{equation}\label{eq:dePows}
\begin{gathered}
\delta P_{1}(t)=-\dot{\epsilon}_{1}(t)[x(t)-\ev{x(t)}],\\
\delta P_{2}(t)=-\dot{\epsilon}_{2}(t)[p(t)-\ev{p(t)}],
\end{gathered}
\end{equation}
in terms of which power fluctuations can be written as
\begin{equation}\label{eq:meanfluct}
D_i (t)=\frac{1}{t-t_0}\int\limits_{t_0}^{t}\mathrm{d}t_2 \int\limits_{t_0}^{t}\mathrm{d}t_1 \ev{\delta P_i (t_2) \delta P_i (t_1)},
\end{equation}
where $t-t_0$ is the total time interval with $t_0\to-\infty$. In this case one can perform the large interval limit  $t-t_0\to\infty$ reducing  the integrated  autocorrelation function to a single integral 

\begin{equation}
D_i (t)=\int\limits_{-\infty}^{t}\mathrm{d}t_1 \left[\ev{\delta P_i (t) \delta P_i (t_1)}+\ev{\delta P_i (t_1) \delta P_i (t)}\right].
\end{equation}
Performing the change of variable $\tau=t-t_1$, one obtains
\begin{equation}
D_i(t)=\int\limits_{0}^{\infty}\mathrm{d}\tau \left[\ev{\delta P_i (t) \delta P_i (t-\tau)}+\ev{\delta P_i (t-\tau) \delta P_i (t)}\right].
\end{equation}

For sake of brevity, in what follows we focus on $D_{1}(t)$, the equations for $i=2$ being analogous. Inserting the expression \eqref{eq:dePows} it follows   
\begin{equation}\label{eq:fluclongt1}
D_1 (t)= \dot{\epsilon}_{1}(t) \int\limits_0^{\infty}\mathrm{d}\tau \dot{\epsilon}_{1}(t -\tau)C(t,t-\tau),
\end{equation}
where 
\begin{equation}
\label{C}
C(t,t')=\ev{x(t) x(t')}+\ev{x(t') x(t)}-2\ev{x(t)}\ev{x(t')}.
\end{equation}
We now demonstrate that from the linearity of Eq. (\ref{eq:diffqoper}), the correlator  $C(t,t')$ can be exactly computed and it does not depend on the external fields $\epsilon_{1/2}$. Moreover, it  reduces to a function of only the difference of time. 

To evaluate Eq.~(\ref{C})  we  exploit the exact  EOM of  $x(t)$  in Eq. (\ref{eq:diffqoper}), still at operatorial level, and  passing in Fourier variables. We have 
\begin{equation}
x(\omega)=\langle x(\omega)\rangle -\chi_{11}(\omega)\xi(\omega).
\end{equation}
Recalling that the fluctuating force has $\langle\xi\rangle=0$ we obtain
\begin{equation}
\langle x(\omega)x(\omega')\rangle=\langle x(\omega)\rangle\langle x(\omega')\rangle +\chi_{11}(\omega)\chi_{11}(\omega')\langle\xi(\omega)\xi(\omega')\rangle.
\end{equation}
The last average is directly computed as Fourier transform of the noise correlator (\ref{xicorr}). We have
\begin{equation}
\langle\xi(\omega)\xi(\omega')\rangle=2\pi m\omega\gamma'(\omega)(1+\coth(\omega/2T))\delta(\omega+\omega').
\end{equation}
Inserting these results in the correlator (\ref{C}) one gets
\begin{equation}
\label{C1}
C(t,t')=C(t-t')= \int_{-\infty}^{\infty}\frac{\mathrm{d}\omega}{2\pi} e^{-i\omega(t-t')}C(\omega),
\end{equation}
with
\begin{equation}
C(\omega)=-2\Im[\chi_{11}(\omega)]\coth(\omega/2T),
\end{equation}
an even function of $\omega$.
We are now in the position to determine the exact expression of the average power fluctuation \eqref{eq:fluclongt1}$, D_1=\frac{1}{\mathcal{T}}\int_{\bar t}^{\mathcal{T}+{\bar t}}\mathrm{d}t D_1 (t)$.
We have
\begin{equation}\label{eq:fluclongt3}
D_1=\frac{\omega^2\epsilon^2_1}{2}\int\limits_{0}^{\infty}\mathrm{d}\tau \cos(\omega\tau)C(\tau)=\frac{\omega^2\epsilon^2_1}{4} C(\omega).
\end{equation}
Inserting Eq.(\ref{C1}) and using the expression \eqref{eq:Onsager} for the Onsager Coefficient $\mathcal{L}_{11}(\omega)$ it follows that 
\begin{equation}\label{eq:fluctfin}
D_1=\epsilon^2_1 \omega\coth(\omega/2T) \mathcal{L}_{11}(\omega).
\end{equation}
It is worth to stress that the above result is exact and valid for any strength of $\epsilon_1$, in other words it does not contain higher order contribution in $\epsilon_1$.
Analogous calculations for the fluctuation $D_2$ give the following result
\begin{equation}\label{eq:fluctfin}
D_2=\epsilon^2_2 \omega\coth(\omega/2T) \mathcal{L}_{22}(\omega).
\end{equation}
Notice that, differently from the average powers in Eq.~\eqref{eq:sol}, which  take the same form in the quantum and classical setting, the power fluctuations substantially differ. In the high temperature regime,~\ie~$T>>\omega$, from Eq.~\eqref{eq:fluctfin} the classical limit follows as       
\begin{equation}\label{eq:fluctclass}
\begin{gathered}
D_1=\epsilon^2_1 2 T \mathcal{L}_{11}(\omega),\\
D_2=\epsilon^2_2 2 T \mathcal{L}_{22}(\omega).
\end{gathered}
\end{equation}

\section{Figures of merit at maximum efficiency}
Here we quote the expressions for the various figures of merit characterizing the isothermal engine, evaluated at the maximum efficiency (ME). 
Recalling the definition $\eta=P_{{\rm out}}/P_{{\rm in}}$, and considering $P_{{\rm out}}=-P_1$ with $P_1<0$ the power associated to the load, and $P_{{\rm in}}=P_2$ with $P_2>0$ the input power,  the efficiency reads
\begin{equation}
\label{etadef}
\eta=-\frac{\epsilon_1^2 {\cal L}_{11} + \epsilon_1\epsilon_2 {\cal L}_{12}}{\epsilon_1\epsilon_2 {\cal L}_{21} + \epsilon_2^2 {\cal L}_{22}}~,
\end{equation}
with Onsager coefficients reported in Eq.~\eqref{eq:Onsager}.
As stated in the main text, we chose the working point by maximizing the efficiency over $\epsilon_1$,~\ie~$\eta_{{\rm  \scriptscriptstyle ME}}$ is defined by ${\rm max}_{\epsilon_1}[\eta]$.
Looking for the maximum over $\epsilon_1$, one gets \cite{BENENTI20171}
\begin{equation}\label{eq:eps1me}
\epsilon_{1,{\rm \scriptscriptstyle ME}}= \epsilon_2 \frac{{\cal L}_{22}}{{\cal L}_{21}}\qty(\frac{1}{\sqrt{1+Y}}-1),
\end{equation}
where
\begin{equation}\label{eq:Y}
Y=\frac{{\cal L}_{12}{\cal L}_{21}}{{\cal L}_{11}{\cal L}_{22} - {\cal L}_{12}{\cal L}_{21}}.
\end{equation}
Inserting Eq.~\eqref{eq:Y} into Eq.~\eqref{etadef} one has
\begin{equation}\label{eq:etame}
\eta_{{\rm \scriptscriptstyle ME}}= X\frac{\sqrt{1+Y}-1}{\sqrt{1+Y}+1},
\end{equation}
where 
\begin{equation}\label{eq:asymm}
X=\frac{{\cal L}_{12}}{{\cal L}_{21}},
\end{equation}
is the asymmetry parameter. 
The corresponding average output power, at ME can be compactly written as
\begin{equation}\label{eq:poutme}
P_{{\rm out, \scriptscriptstyle ME}}= \epsilon_2^2 \eta_{{\rm \scriptscriptstyle ME}}\frac{{\cal L}_{22}}{\sqrt{1+Y}},
\end{equation}
with associated power fluctuation
\begin{equation}\label{eq:Doutme}
D_{{\rm out, \scriptscriptstyle ME}}= \epsilon^2_{1,{\rm \scriptscriptstyle ME}} \omega \coth(\omega/(2T)){\cal L}_{11}.
\end{equation}

Note that the above expressions enter into the definition of the relative fluctuations, depicted in Fig.~2(c) in the main text
\begin{equation}\label{eq:sigmame}
\Sigma_{{\rm \scriptscriptstyle ME}}=\sqrt{\frac{D_{{\rm out,\scriptscriptstyle ME}}}{P^2_{{\rm out,\scriptscriptstyle ME}}}}.
\end{equation}
%From the previous expressions and from Eq.\eqref{eq:Onsager}, it is evident that the conversion performance can be evaluated in any parameter regime. 
\subsection{Power and efficiency with broken TRS}
\begin{figure}[h]
	\centering
	\includegraphics[width=0.95\linewidth]{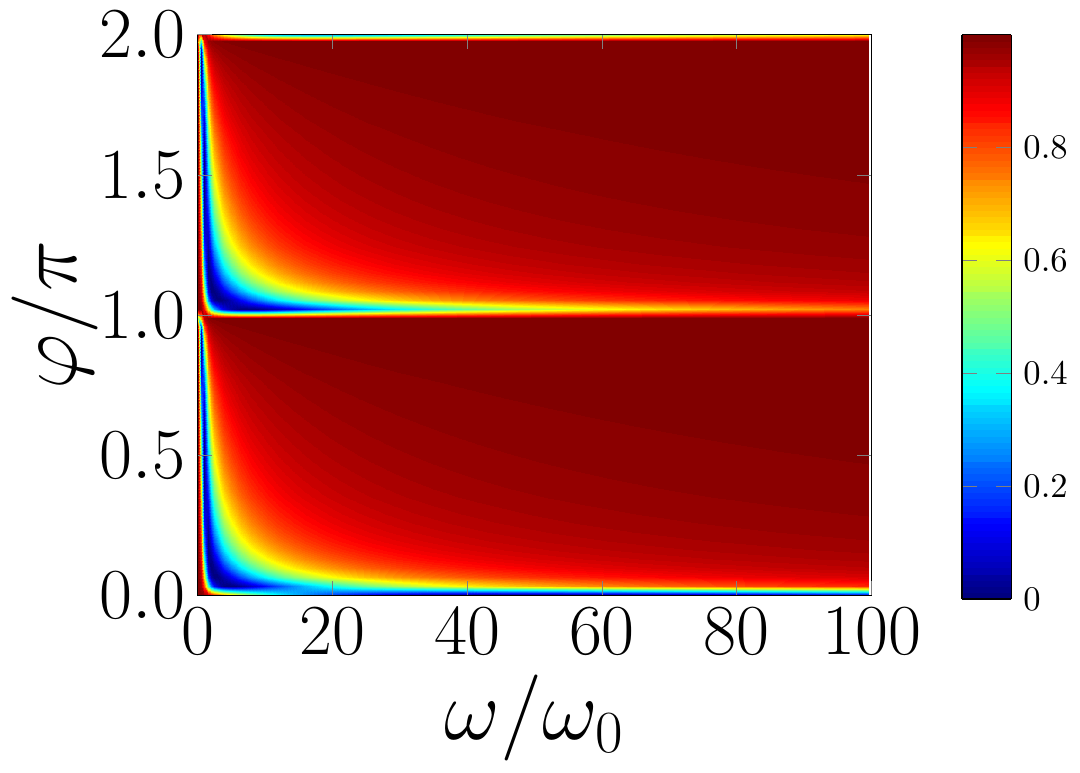}
	\caption{Density plot of the maximum efficiency $\eta_{\rm{\scriptscriptstyle ME}}$ as function of $(\omega,\varphi)$ for $s=1$, $m\epsilon^2_2=0.025\text{}\omega_{0}$,$\gamma_{\rm{1}}=0.5\text{}\omega_{0}$, and $\bar{\omega}=\omega_0$.}
	\label{fig:densityeta} 
\end{figure}

\begin{figure}[h]
	\centering
	\includegraphics[width=0.95\linewidth]{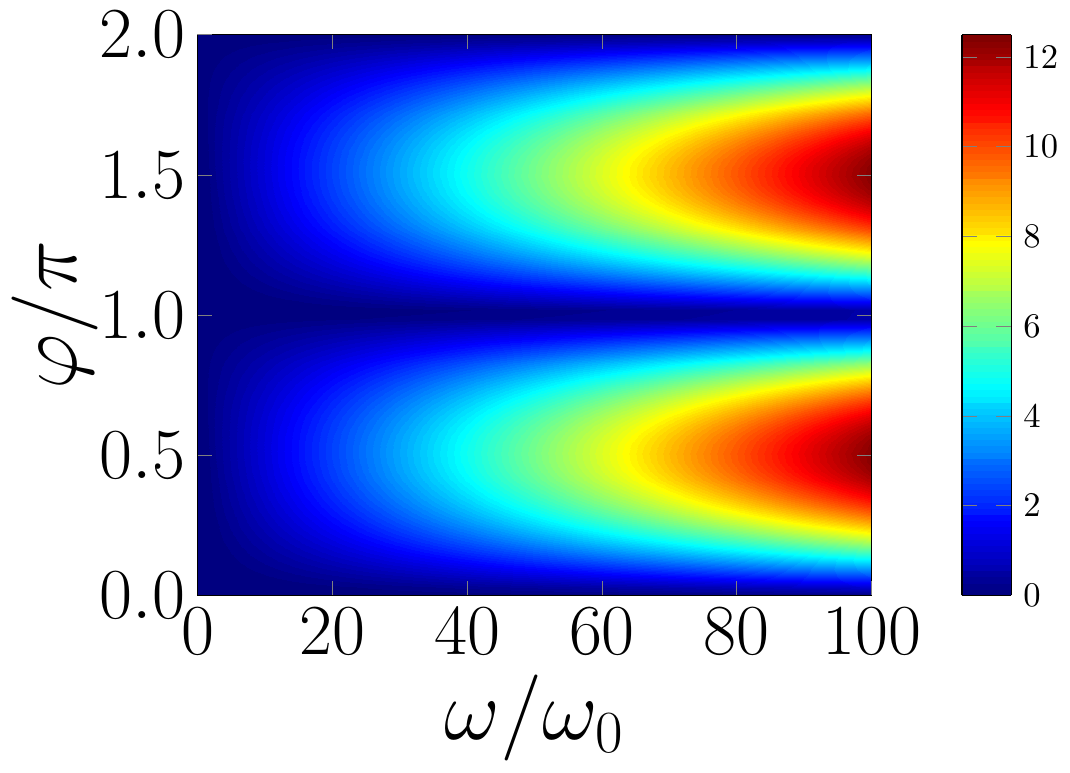}
	\caption{Density plot of the output power at ME $P_{\rm{out, ME}}/\omega^2_{0}$ as function of $(\omega,\varphi)$ for $s=1$, $m\epsilon^2_2=0.025\text{}\omega_{0}$,$\gamma_{\rm{1}}=0.5\text{}\omega_{0}$, and $\bar{\omega}=\omega_0$.}
	\label{fig:densitypow} 
\end{figure}

The asymmetry parameter written in Eq.~\eqref{eq:asymm} is related to the breaking of time reversal symmetry (TRS). Indeed, for $\varphi=0$ or $\pi$ one has $X=1$ (symmetric case), while other values of $\varphi$ would break TRS with $X\neq 1$.
In the main text we have shown results fixing the phase factor $\varphi=3\pi/2$ where $X=-1$,~\ie~the Onsager matrix is completely asymmetric. 
In the following we will focus the discussion by considering the limit of large cut-off frequency $\omega_c\gg\omega$. In Fig.~\ref{fig:densityeta} we report the density plot of the maximum efficiency $\eta_{\rm{\scriptscriptstyle ME}}$ as a function of $(\omega,\varphi)$, for $s=1$ and at fixed field amplitudes and dissipation strength. 
A similar plot is reported also in Fig. \ref{fig:densitypow} for the output power at ME. 
We observe that at fixed driving frequency $\omega$ both quantities are periodic functions of the  phase difference $\varphi$, with period $\pi$. In addition one can see that setting the engine operating regime at maximum asymmetry,~\ie~$\varphi=k\pi/2, k\,\text {odd}$, an efficiency near to 1 with a finite optimal output power can be achieved in the anti-adiabatic regime of   high frequencies.


\subsection{Quasi-static and anti-adiabatic limits}
It is useful to report the behavior of the quantities in Eq.~\eqref{eq:etame}-\eqref{eq:sigmame}, for $0< s < 2$, in the two opposite frequency regimes: quasi-static limit ($\omega \to 0$) and anti-adiabatic (high frequencies, see below) regime.
These asymptotic behaviours can be evaluated analytically for generic dissipation with power-law $s$, starting from the explicit expressions of the Onsager coefficients reported in Eq.~\eqref{eq:Onsager} and in the main text.~We first discuss the  low frequency regime.~For sake of simplicity, we fix $\varphi=3\pi/2$.~From Eq.~\eqref{eq:Onsager}, the leading terms in the limit $\omega\to 0$ are
\begin{eqnarray} 
  \mathcal{L}_{11}(\omega)&\simeq&\frac{\bar{\gamma}_{{s}}}{2 m} \frac{\omega^{s+1}}{\omega^4_{0}},\nonumber\\ 
\mathcal{L}_{12}(\omega)&=&-\mathcal{L}_{21}(\omega)\simeq-\frac{1}{2}\frac{\omega^2}{\omega^2_{0}},\\
\mathcal{L}_{22}(\omega)&\simeq& \frac{m\bar{\gamma}_{{s}}}{2}\frac{\omega^{s+3}}{\omega^4_{0}},\nonumber
\end{eqnarray}
with $\bar\gamma_s=\gamma_s \bar\omega^{1-s}$.
This implies that for $\omega \to 0$ one obtains
\begin{eqnarray}
&&\eta_{{\rm \scriptscriptstyle ME}}\simeq 1-2 \frac{\gamma_s\bar{\omega}}{{\omega_0^2}} \qty(\frac{\omega}{\bar{\omega}})^s,\nonumber\\
&&P_{{\rm out,\scriptscriptstyle ME}}\simeq \epsilon_2^2 \frac{m\omega^3}{2\omega_0^2},\nonumber\\
&&D_{{\rm out, \scriptscriptstyle ME}}\simeq \frac{\gamma_{\rm{s}} m \epsilon^2_2 \bar{\omega}^5}{2 \omega_{0}^4} \coth(\frac{\omega}{2T})  \qty(\frac{\omega}{\bar{\omega}})^{s+4},\nonumber\\
&&\Sigma_{{\rm \scriptscriptstyle ME}}\simeq \sqrt{\frac{2 \gamma_{\rm{s}}}{m \epsilon^2_2 \bar{\omega}} \coth(\frac{\omega}{2T})\qty(\frac{\omega}{\bar{\omega}})^{s-2}}\,.
\end{eqnarray}
Notice that, although in the quasi-static regime the efficiency tends to the Carnot limit, the output power and the power fluctuations vanish, while the relative fluctuations diverge.~The memory effects enter only the exponents of the different figures of merit.~It follows that in the quasi-static regime energy conversion performance close to Carnot limit can be only achieved with vanishing output power, as recently discussed in related literature \cite{Shiraishi:untradeoff}.  

Let us consider now the opposite anti-adiabatic regime at large frequencies defined by the condition $\omega\gg \tilde\omega$ with $\tilde\omega=\text{max }\qty{\omega_0,\bar{\omega}(\gamma_s/\bar{\omega})^{1/(2-s)}\qty[1+\cot^2(\pi s/2)]^{1/(4-2s)}}$. In this regime,
 the following scaling for the figures of merit are obtained
\begin{eqnarray}\label{eq:54}
&&\eta_{{\rm \scriptscriptstyle ME}}\approx 1 - 2 \frac{\gamma_{\rm s}}{\bar{\omega}} \qty(\frac{\omega}{\bar{\omega}})^{s-2},\nonumber\\
&&P_{{\rm out, \scriptscriptstyle ME}}\approx \frac{\epsilon_2^2m}{2}\omega,\nonumber\\
&&D_{{\rm out,\scriptscriptstyle ME}}\approx \frac{m}{2}\epsilon_2^2\gamma_{\rm s}\bar{\omega}  \coth(\frac{\omega}{2T})\qty(\frac{\omega}{\bar{\omega}})^s,\nonumber\\ 
&&\Sigma_{\rm{\scriptscriptstyle ME}}\approx \sqrt{\frac{2 \gamma_{\rm s}}{m\epsilon_2^2 \bar{\omega}}\coth(\frac{\omega}{2T})\qty(\frac{\omega}{\bar{\omega}})^{s-2}}.
\end{eqnarray}

From the above asymptotic scalings, one can recognize that also in the anti-adiabatic regime the maximum efficiency tends to the Carnot limit $\eta_{{\rm ME}}\to 1$.~However, while in the low frequency regime the output power vanishes $\propto \omega^3$, here a diverging power $\propto \omega$ is obtained. Furthermore, at high frequency, relative uncertainties tend to vanish, thus the isothermal engine shows remarkable performance.~It is worth to note  that memory effects, related to non-Ohmic dissipation, do not affect the behavior of the leading term in the output power.~An explicit dependence on $s$ enters only into sub-leading contributions, which scale as $O(\omega^{s-1})$.

\section{Trade-off parameter and bounds}
In the context of thermodynamic uncertainty relations (TURs) it is often introduced a so-called trade-off parameter
\begin{equation}
{\cal Q}=\sigma \frac{D_{{\rm out}}}{P^2_{{\rm out}}}= \frac{1}{T}(P_{{\rm in}}-P_{{\rm out}})\frac{D_{{\rm out}}}{P^2_{{\rm out}}},
\end{equation}
where $\sigma=(P_{{\rm in}}-P_{{\rm out}})/T$ is the entropy production rate. This parameter quantifies the ability to achieve simultaneously good engine performance with finite output power and, possibly, low power fluctuations.
As mentioned in the main text, several bounds have been introduced constraining the minimum value of the trade-off parameter.~In particular, Koyuk and Seifert\cite{Koyuk:BoundPer}, considering a system under time-dependent drives, have derived the bound ${\cal Q}\geq {\cal V}$ with
\begin{equation}\label{eq:Seifbound}
\mathcal{V}=2 \qty(1-\omega\frac{\partial_\omega P_{{\rm out}}}{P_{{\rm out}}})^2.
\end{equation}
Notice that this relation generalizes previous ones valid only for static driving \cite{Pietzonka:Tradeoff}, and holds true assuming overdamped and Markovian dynamics.

%\textcolor{red}{ATTENZIONE !! SIAMO SICURI CHE NON CI SIANO ALTRE ASSUNZIONI IN SEIFERT RELATIVE A BILANCIO DETTAGLIATO NEI RATE AD OGNI ISTANTE??}

Considering the working point of the isothermal engine at maximum efficiency, the trade-off parameter can be written as
\begin{equation}
{\mathcal Q}_{{\rm \scriptscriptstyle ME}}=\frac{1}{T}\qty(\frac{1}{\eta_{{\rm \scriptscriptstyle ME}}}-1)\frac{D_{{\rm out, \scriptscriptstyle ME}}}{P_{{\rm out, \scriptscriptstyle ME}}}.
\end{equation}

%\textcolor{red}{IO METTEREI LA SCALA SULLE X PIU' STRETTA PARTENDO DA $10^{-2}$ A $10^{-2}$, MI SEMBRA CHE ORA 
%SIA ANCORA PIU' LARGA}

\begin{figure}[h]
	\centering
	\includegraphics[width=0.97\linewidth]{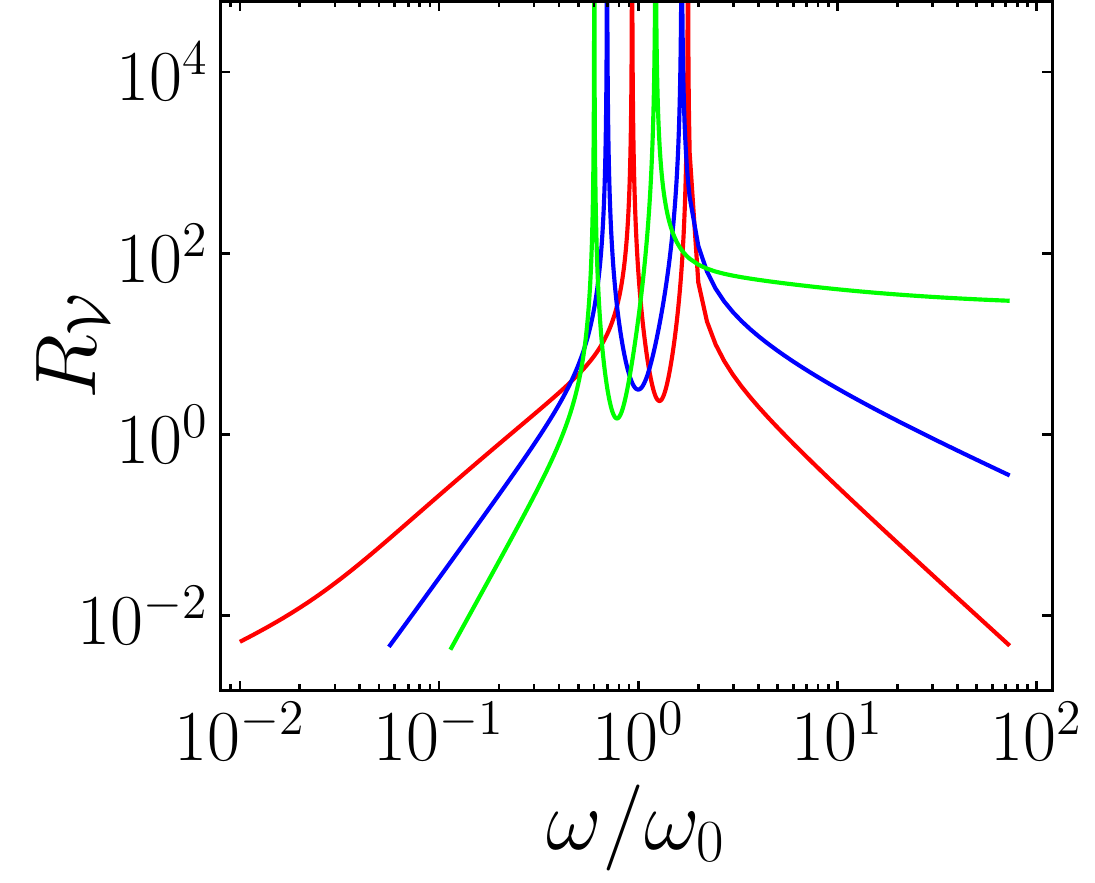}
	\caption{Ratio of the trade-off parameter and the Seifert bound $\mathcal{V}$ at maximum efficiency plotted against the driving frequency $\omega$ for $s=\{0.5,1,1.5\}$ (red line, blue line, green line respectively). The model parameters have been set as follows: \textcolor{black}{$m\epsilon^2_2=0.025\text{}\omega_{0}$}, $\gamma_{\rm{s}}=0.5\text{}\omega_{0}$, $\varphi=3\pi/2$, $T=0.01\text{}\omega_{0}$, and $\bar{\omega}=\omega_0$.}
	\label{fig:seiffig} 
\end{figure} 
In Fig.~\ref{fig:seiffig}, we show the ratio of the trade-off parameter and the bound in Eq.~\eqref{eq:Seifbound}, $R_{\scriptscriptstyle \mathcal{V}}=({\cal Q}/\mathcal{V})_{{\rm \scriptscriptstyle ME}}$, computed at ME, for different kinds of dissipation,~\ie~$s=0.5,1,1.5$.~It is evident that the engine performance violates the bound in Eq.~\eqref{eq:Seifbound}, since several regions in the quasi-static and anti-adiabatic regimes exist where the ratio $R_{\scriptscriptstyle \mathcal{V}}$ falls below 1.~
%\textcolor{red}{parametri giusti di s?}
Indeed, the violation is linked to the behavior of ${\cal Q}_{{\rm \scriptscriptstyle ME}}$ in the two opposite limits.
%It is worth noticing that this bound is violated both in the quasi-static ($\omega\ll \gamma_{\rm s},\omega_0$) and in the anti-adiabatic regime ($\omega\gg \gamma_{\rm s},\omega_0$), as shown in Fig.\ref{fig:seifbound}.
At high frequencies,~\ie~$\omega\gg \tilde{\omega},T$ one gets the following asymptotic behaviors
\begin{equation}\label{eq:trdoffanti}
%{\cal Q}_{{\rm ME}}\approx \frac{2}{T} \bar{\gamma}_{\rm s}^2  \coth(\omega/(2T)) \omega^{2s-3}
{\mathcal Q}_{{\rm \scriptscriptstyle ME}}\approx \frac{2 \gamma^2_{\rm s}}{T\bar{\omega}}  \qty(\frac{\omega}{\bar{\omega}})^{2s-3},
\end{equation}
and
\begin{equation}
{\cal V}_{{\rm \scriptscriptstyle ME}}\approx 2,
\end{equation}
which shows that the ratio $({\mathcal Q}/{\mathcal V})_{{\rm \scriptscriptstyle ME}}$ tends to vanish for $0<s<3/2$, while the bound is not violated for $s>3/2$.
\begin{figure}[htb]
	\centering
	\includegraphics[width=0.97\linewidth]{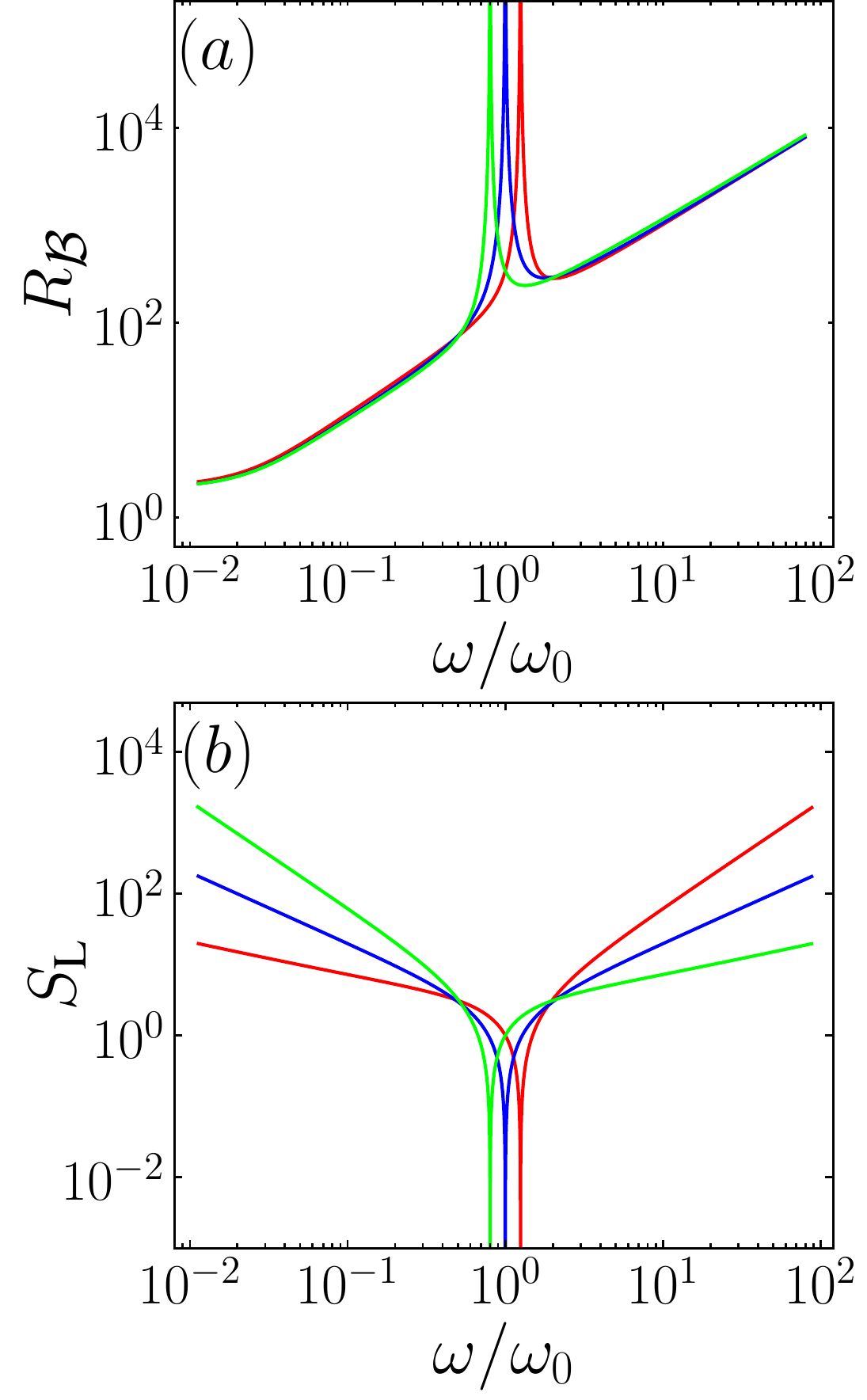}
	\caption{Comparison of the engine performance against the unified TUR bound \textcolor{black}{$\mathcal{B}$}.~In Panel $(a)$, the ratio \textcolor{black}{$R_{\scriptscriptstyle \mathcal{B}}=(\mathcal{Q}/\mathcal{B})_{{\rm  \scriptscriptstyle ME}}$} is plotted as function of the driving frequency for $s=\{0.5,1.0,1.5\}$ (red line, blue line, green line respectively).~In Panel $(b)$, the asymmetry factor $S_{\rm L}$ at ME is plotted versus the driving frequency for the same values of the parameter $s$. In both panels, the model parameters have been set as follows: $m\epsilon^2_2=0.025\text{}\omega_{0}$, $\gamma_{\rm{s}}=0.5\text{}\omega_{0}$, $T=0.01\text{}\omega_{0}$, $\varphi=3\pi/2$, and $\bar{\omega}=\omega_0$.}
	\label{fig:Brandner} 
\end{figure} 

Although in the two opposite regimes we can observe violation of this bound, in this case the asymptotic power law is different.~Indeed, for $\omega\to 0$ one has
\begin{equation}
{\mathcal Q}_{{\rm \scriptscriptstyle ME}}\approx \left(\frac{2 \gamma^2_{\rm s} \bar{\omega} }{\omega_0^2}\right)^2\qty(\frac{\omega}{\bar{\omega}})^{2s},
\end{equation}
and
\begin{equation}
{\mathcal V}_{{\rm \scriptscriptstyle ME}}\approx 2.
\end{equation}

It is found that the bound in Eq.~\eqref{eq:Seifbound} is never obeyed in the quasi-static regime. We note that these possible violations in the quasi static regime are peculiar features of the model considered in this work. In particular, it is linked to the fact that one of the two time-dependent drives is coupled to the momentum $p$.

%\textcolor{red}{ATTENZIONE NON SONO SICURA CHE LA SPIEGAZIONE CHE SEGUE RELATIVA AL FATTO CHE NON SIAMO MAI OVERDAMPED SIA SIA GIUSTA!}

Indeed, from the associated equation of motion (see Eq.~\eqref{eq:EOM}), we note that this model presents a fully underdamped dynamics in the whole frequency regime and, even in the quasi-static limit, it {\it does not} collapse to an overdamped motion for finite $\epsilon_2$. %which would fulfill the bound in Eq.~\eqref{eq:Seifbound} for $s=1$ (Ohmic dissipation and Markovian dynamics).
Following these results, it is thus interesting to look for different formulations of the TUR bounds which could describe heat engine performance in the peculiar regime such as the one investigated here.   

Under very general conditions, by making the assumption of validity of the linear response regime, Macieszczak et al.
 \cite{Brandner:unifiedTUR} have derived a unified TUR bound. The bound states that
\begin{equation}
\label{eq:bound}
{\mathcal Q}\geq {\mathcal B}=\frac{2}{1 + S^2_\text{L}}~,
\end{equation}
where $S_{\rm L}$ indicates the asymmetry of the Onsager matrix \cite{Brandner:unifiedTUR}.
The model of isothermal engine under investigation, although not restricted to small values of the external driving strength, is written in terms of generalized Onsager coefficients and fullfill the bound posed by Eq.~\eqref{eq:bound} in the whole frequency range (see Fig.~\ref{fig:Brandner}, panel $(a)$).

Here, we report  the frequency dependence of $S_{\rm L}$, that in the fully asymmetric case $\varphi=3\pi/2$ can be compactly written as \cite{Brandner:unifiedTUR} 

\begin{equation}
\label{eq:slbrand}
S_{\rm L}=|\mathcal{L}_{12}|/\sqrt{\mathcal{L}_{11}\mathcal{L}_{22}}.
\end{equation}
It is reported in Fig.~\ref{fig:Brandner}(b), where in the two investigated frequency regions, one can notice a diverging behaviour with frequency, occurring with different power-laws, for various $0 <s<2$. This behaviour agrees with the linear scaling, at high frequency, obtained for the ratio $R_{\scriptscriptstyle \mathcal{B}}=({\mathcal Q}/{\mathcal B})_{{\rm \scriptscriptstyle ME}} \propto \omega$.
    
Finally, it is worth to report also the scaling behavior at high frequency of the entropy production at maximum efficiency,~\ie~$\sigma_{{\rm \scriptscriptstyle ME}}$.~From Eq.\eqref{eq:54} it reads
\begin{equation}
\sigma_{{\rm \scriptscriptstyle ME}}\approx \frac{\epsilon^2_{2} m \gamma_{s}}{T} \abs{\frac{\omega}{\bar{\omega}}}^{s-1}.
\end{equation}
One can thus observe that, in the anti-adiabatic regime,  sub-Ohmic bath ($s<1$) results in a vanishing entropy production, while the latter saturates to a constant value in the Ohmic case($s=1$).
Moreover, considering super-Ohmic bath ($1<s<2$), the entropy production tends to grow with power law $\propto \omega^{s-1}$, whereas the squared relative power fluctuations keep on decreasing with power $\propto \omega^{s-2}$. As a consequence, the tradeoff parameter behaves as in Eq.\eqref{eq:trdoffanti}.

\bibliography{supp}